# Diagnosis of Knee Osteoarthritis Using Bioimpedance & Deep Learning

Jamal Al-Nabulsi
*Department of Medical Engineering*
Al-Ahliyya Amman University
Ammaan, Jordan
j.nabulsi@ammanu.edu.jo

Mohammad Al-Sayed Ahmad
*Department of Medical Engineeringg*
Al-Ahliyya Amman University
Ammaan, Jordan
20191169@ammanu.edu.jo

Baraa Hasaneiah
*Department of Medical Engineeringg*
Al-Ahliyya Amman University
Ammaan, Jordan
202010511@ammanu.edu.jo

Fayhaa AlZoubi
*Department of Medical Engineeringg*
Al-Ahliyya Amman University
Ammaan, Jordan
202010177@ammanu.edu.jo

*Abstract*— **Diagnosing knee osteoarthritis (OA) early is crucial for managing symptoms and preventing further joint damage, ultimately improving patient outcomes and quality of life. In this paper, a bioimpedance-based diagnostic tool that combines precise hardware and deep learning for effective non-invasive diagnosis is proposed. system features a relay-based circuit and strategically placed electrodes to capture comprehensive bioimpedance data. The data is processed by a neural network model, which has been optimized using convolutional layers, dropout regularization, and the Adam optimizer. This approach achieves a 98% test accuracy, making it a promising tool for detecting knee osteoarthritis and potentially other musculoskeletal disorders.**

Keywords— Deep Learning, Knee Osteoarthritis, Artificial Intelligence, Bioimpedance.

## I. INTRODUCTION

Osteoarthritis is a common and debilitating condition marked by the degeneration of joint cartilage and underlying bone, resulting in pain, stiffness, and reduced mobility. Among the affected joints, the knee is the most frequently impacted by OA. This joint disease leads to changes such as worn cartilage, bone spurs, reduced joint spacing, and swelling [1]. Similarly, osteoarthritis of the hip is recognized as one of the most disabling and prevalent dysfunctions in the population [2]. Diagnosis and monitoring of OA progression are typically conducted using joint imaging techniques like X-ray or ultrasound, alongside self-reports of OA-related symptoms such as pain, fatigue, and mobility issues. While imaging methods can detect changes in joint spacing and hard tissues, they are limited to late-stage OA and require repeated clinical visits over extended periods, placing a significant burden on patients and demanding substantial resources in terms of equipment and medical personnel. Although these imaging-based techniques provide accurate assessments of knee conditions, they necessitate expensive, large-scale infrastructures and the expertise of professionally trained medical staff [1,3-5]. Electrical bioimpedance is a method that involves sending a small electrical current through a volume of biological tissue and then measuring the resulting voltage change across that tissue [6]. This measurement is used to calculate the passive impedance that the tissue presents against the flow of electrical current, Which gives a considerable variation in electrical conductivity among different tissue types. Electrical bioimpedance measurements can yield valuable insights into the structural composition of the tissue [7,8]. A study on individuals with knee osteoarthritis utilizing bioimpedance spectroscopy found that impedance values tend to increase in correlation with the severity of the disease [9]. Early detection of osteoarthritis can facilitate the prompt initiation of interventions and therapeutic strategies [10]. Osteoarthritis is highly prevalent worldwide, with an increase of 113.25% from 1990 to 2019 [1]. In instances of severe joint degradation, arthroplasty may be required, leading to significant healthcare costs [11]. Reducing 5 kg from the standard BMI range in overweight and obese individuals can significantly help prevent 24% of knee osteoarthritis surgical cases. Additionally, early detection of arthritis can also help avoid the need for surgery [12].

In [13], authors used a bipolar electrode configuration to examine bioimpedance measurement for diagnosing knee injuries. It was reported in this work to place electrodes on acupuncture points near the knee to simplify the process compared to the more complex tetrapolar setup. This approach involved measuring bioimpedance across a frequency range of 100 Hz to 1 MHz, focusing on changes associated with knee injuries, particularly fluid accumulation. Controlled laboratory tests on healthy volunteers assessed external factors, followed by evaluations on both healthy and injured knees. Analysis using Principal Component Analysis (PCA) and Support Vector Machines (SVM) revealed a significant reduction in bioimpedance in injured knees around 100 kHz, correlating with fluid buildup. This method achieved a sensitivity of 87.18% in distinguishing between healthy and injured knees, demonstrating its practical and reliable application for point-of-care diagnostics. Similarly, utilized bioimpedance spectroscopy to develop a new index for non-invasive assessment of knee osteoarthritis in Brazilian military parachuters, involving twelve male volunteers aged 20-45 years [14]. Bioimpedance spectroscopy data were collected using a prototype device and processed to extract resistive and reactive components at 50 kHz. Statistical methods differentiated healthy from osteoarthritic knees, with significant differences in bioimpedance parameters and a strong correlation with the

Dejour classification system, validating the new index despite the study's small sample size. This suggests that bioimpedance spectroscopy could be a viable tool for early osteoarthritis diagnosis and monitoring, especially in high-impact activity populations. Meanwhile, in [15], a cross-sectional study to assess bioelectrical impedance parameters in individuals with hip osteoarthritis compared to healthy was conducted. This study involved 54 participants aged 45-70, with 31 in the OA group and 29 in the control group. Key bioimpedance parameters such as phase angle, impedance, reactance, and muscle mass were measured using segmental bioelectrical impedance analysis at 50 kHz, revealing significant differences between affected and contralateral limbs in the OA group, including a decrease in phase angle and muscle mass and an increase in impedance. Additionally, [10] aimed to validate bioelectrical impedance plethysmography as a cost-effective diagnostic tool for knee joint osteoarthritis through animal studies, a pilot human study, and a clinical investigation. This involved impedance measurements in anesthetized albino rats' knee joints, freshly amputated bull calves' knee joints before and after saline injections, and human subjects under static and dynamic loading conditions. Significant differences in impedance measurements between normal and osteoarthritic knee joints across all phases indicated that bioelectrical impedance plethysmography effectively distinguishes pathological changes in the knee joint, suggesting it as a viable, non-invasive alternative to traditional imaging methods like X-ray and MRI for early detection of OA.

## II. Methodology

### A. Data Set

The dataset comprises impedance measurements taken during various knee exercises, specifically targeting participants at different stages of knee osteoarthritis. It includes readings from 28 individuals, with 13 diagnosed with OA and 15 identified as healthy controls. Each participant's knee impedance readings were meticulously recorded twenty times to ensure comprehensive data collection. These measurements were taken during exercises such as cyclic movement, gait analysis, standing, knee extension, and knee flexion. The dataset, sourced from the IEEE Data Port, is thus rich with detailed impedance profiles reflecting how knee bioimpedance varies across different activities and OA severity levels, providing a robust foundation for the diagnostic model's training and evaluation [16].

### B. Hardware Design

The hardware system design consists of two primary electrical circuits, each serving distinct functions for bioimpedance measurement and sequencing. The first circuit is designed to measure bioimpedance across various knee regions using integrated circuits the AD5933 Impedance Converter IC, which combines a frequency generator with a 12-bit analog-to-digital converter (ADC) for precise complex impedance measurements. This IC excites external impedance at a specific frequency and samples the resulting signal with the onboard ADC. Then the digital signal processor DSP engine performs a discrete Fourier transform (DFT) to produce real and imaginary data outputs, facilitating accurate impedance magnitude and phase calculations. The circuit also incorporates AD8606 dual amplifiers, which provide rail-to-rail input and output with minimal offset voltage, low noise, and a broad signal bandwidth, making them suitable for applications requiring high precision, such as high-impedance sensors. Additionally, the ADP150 Voltage Regulator supplies a stable, low-noise power output, essential for sensitive electronic circuits. The second circuit, known as the sequence circuit, organizes the sequence of impedance readings across eight knee electrodes. Electrode locations are shown in "Fig. 1". These relays, which close and enable readings when voltage is applied, help minimize voltage drop and interference, thus enhancing measurement accuracy. Sequential impedance readings between different electrode pairs are systematically recorded, creating a comprehensive dataset. To manage data acquisition and control, an Arduino Mega microcontroller is employed due to its cost-effectiveness, support for I2C communication, and availability of multiple pins for relay control. The entire hardware setup, including surface-mount device (SMD) components, is mounted on a printed circuit board (PCB) ensuring a layout that minimizes noise and maintains reliable connections between components Overall Hardware design is presented in "Fig. 2".

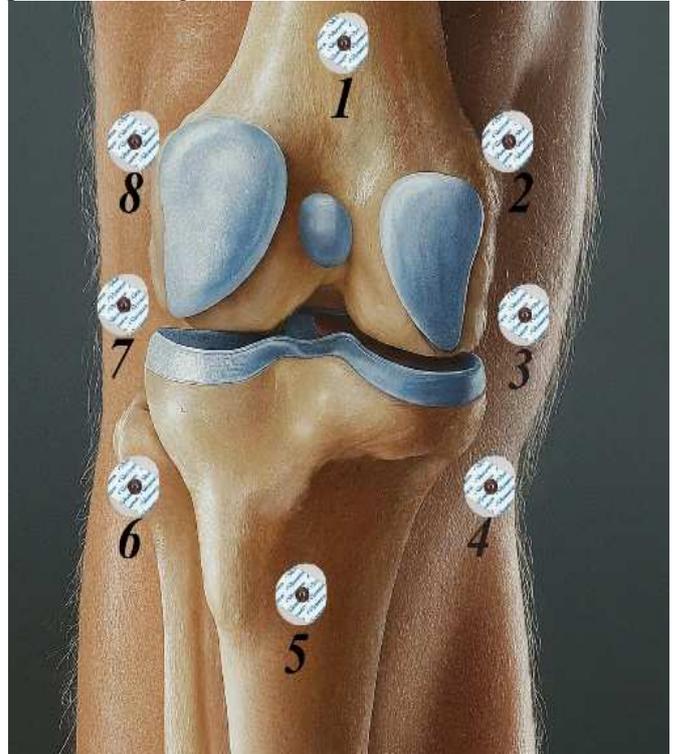

Fig. 1. Electrode locations on the knee

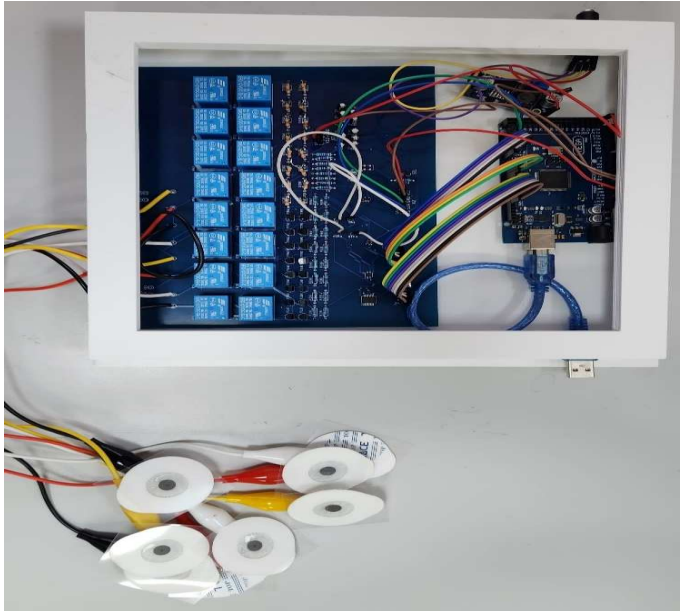

Fig. 2. Hardware design

## C. Software design

The data preprocessing phase starts with loading the dataset from a CSV file that contains bioimpedance readings related to knee osteoarthritis. This dataset includes measurements taken during various exercises, performed by participants at different stages of OA severity. To maintain the integrity and reliability of the data, the first step is to handle any missing values; these are removed to ensure that subsequent analyses are not compromised by incomplete data. Next, categorical variables such as the type of exercise performed, the participant ID, and the electrode pattern used during the measurements need to be converted into numerical formats for further processing. This is achieved using a Label Encoder, which assigns a unique numerical value to each category. For example, different exercises like 'Cyclic', 'Gait', 'Standing', 'Extension', and 'Flexion' are given a distinct numerical code. Similarly, each participant and electrode pattern is encoded numerically. The target variable, which indicates the presence or severity of osteoarthritis (referred to as "affectation"), also undergoes encoding. This ensures that the classification models can effectively process and learn from this data. Through these preprocessing steps, the dataset is transformed into a format suitable for deep learning algorithms, facilitating accurate analysis and prediction of knee osteoarthritis severity. The features and target variable are then separated, and the features are scaled using Standard Scaler to normalize the data. The dataset is redistributed into 90% for training and 10% for testing. The model architecture is sequential, comprising convolutional layers, pooling layers, dropout for regularization and fully connected layers to handle the temporal and spatial features in bioimpedance data. Initially, a `Conv1D` layer with 64 filters and a kernel size of 3 uses ReLU activation to extract local patterns from the bioimpedance readings. This is followed by another Conv1D layer with 128 filters to capture more complex features. Dimensionality is reduced via `MaxPooling1D` with a pool size of 2 after each convolutional layer and dropout layers with rates of 0.5 and 0.6 follow the first and second pooling layers, respectively, to mitigate overfitting. The feature maps are flattened into a 1D vector using a `Flatten` layer, then input into a `Dense` layer with 256 units and ReLU activation for high-level representation learning. An additional dropout layer with a rate of 0.6 further reduces overfitting. The final output layer, a `Dense` layer with SoftMax activation, provides multi-class classification outputs. The model is compiled using the Adam optimizer with a learning rate of 0.000065 to ensure gradual convergence, and categorical cross-entropy loss is employed, suitable for multi-class classification. The model includes accuracy as a performance metric. Training is conducted using the fit method on the training data (X-train, Y-train), spanning 40 epochs with a batch size of 32. Early stopping with a patience of 10 epochs monitors the training loss, halting training when no improvement is observed and restoring the best-performing model weights to prevent overfitting.

## III. RESULTS

The bioimpedance measurement system demonstrated robust performance, with electrodes strategically positioned around the knee "Fig. 1" showing consistent results across various subjects. The hardware setup achieved stable readings with minimal noise interference through precise circuit design and effective shielding of measurement pathways, ensuring accurate impedance data collection. The sequencing circuit efficiently managed switching between electrode pairs, facilitating comprehensive bioimpedance measurements across different knee regions without significant voltage drops, thereby maintaining signal integrity. The collected bioimpedance data were accurately transmitted to the Arduino microcontroller and displayed on the serial monitor, with validated transmission confirming data readiness for further processing. The software results from model training revealed that the `Conv1D` layers with 64 and 128 filters effectively extracted both local and complex features from the bioimpedance data, with the ReLU activation function enhancing the learning of non-linear patterns. The `MaxPooling1D` layers reduced the dimensionality of feature maps, while `Dropout` layers with rates of 0.5 and 0.6 mitigated overfitting by randomly deactivating neurons during training. Continuous improvements in loss reduction and accuracy were observed throughout the training process, reaching about 97% training accuracy and minimal loss, as shown in "Fig. 3"and "Fig. 4". Early stopping, with a patience of 10 epochs, halted training at the optimal point, restoring the best-performing model weights and preventing overfitting, thus retaining the model's best performance on the validation set. Diagnostic accuracy was further demonstrated with a test accuracy of 98%, indicating the model's strong capability in classifying bioimpedance readings for diagnosing knee osteoarthritis. The confusion matrix "Fig. 5" showed high accuracy in distinguishing various osteoarthritis severity levels, with rare misclassifications, underscoring the model's ability to differentiate between OA stages effectively.

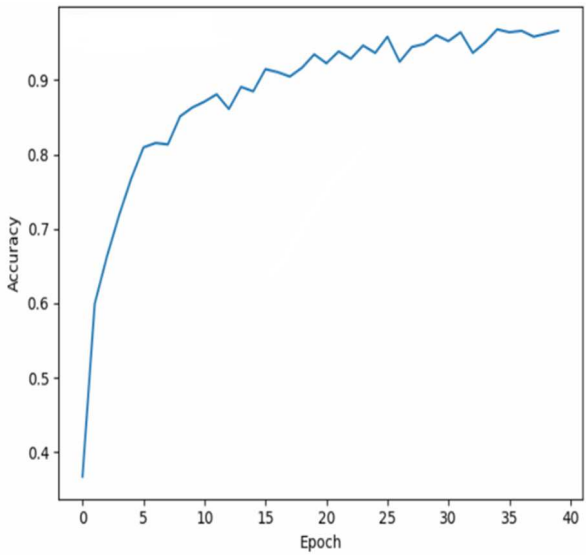

Fig. 3. Training Accuracy

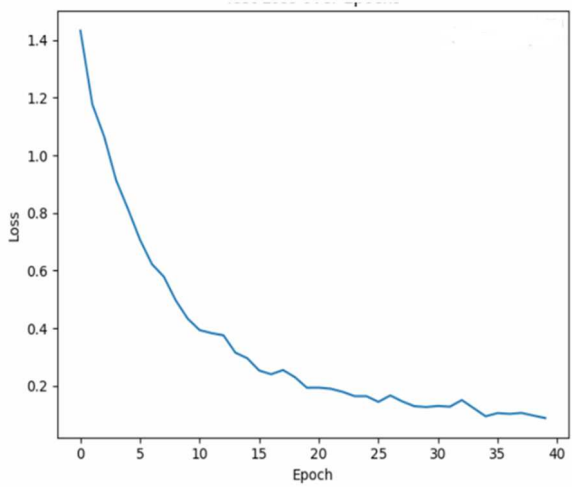

Fig. 4. Training loss

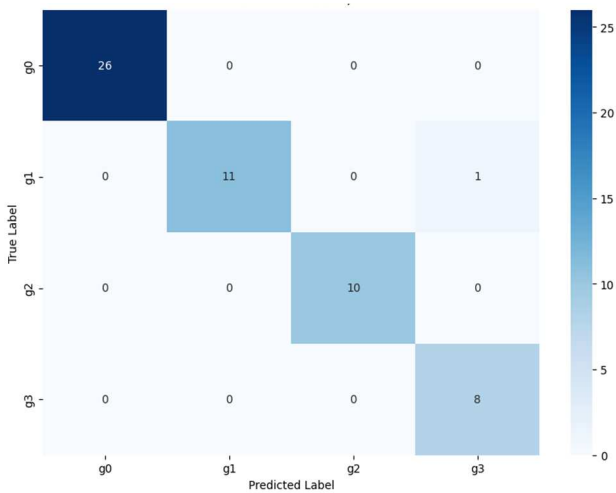

Fig. 5. Confusion Matrix

The classification performance metrics for the neural network model, as detailed in Table 1, indicate its exceptional accuracy in diagnosing knee osteoarthritis across various severity levels. The model demonstrates perfect precision, recall, and F1-scores for OA severity levels g0 and g2, highlighting flawless classification for these groups. For severity level g1, the model maintains perfect precision and a slightly lower recall, resulting in an excellent F1-score. In the case of severity level g3, the model shows high precision and recall, with a minor reduction in precision. The model's overall accuracy is 98%, reflecting its robustness and reliability. Both macro and weighted averages of precision, recall, and F1-scores are consistently high, emphasizing the model's effectiveness in classifying OA severity with minimal error. This performance, as indicated in Table 1, suggests that the neural network model is a promising tool for non-invasive OA diagnostics, capable of accurately distinguishing between different OA severity levels. The Receiver Operating Characteristic (ROC) curve depicted in the image shows the performance of the neural network model in classifying knee osteoarthritis severity levels. "Fig. 6" shows the curve for each class g0, g1, g2, and g3 achieving an Area Under the Curve (AUC) of 1.00, indicating perfect classification performance. The ROC curve plots the true positive rate against the false positive rate, and an AUC of 1.00 signifies that the model has achieved flawless discrimination between all OA .

TABLE I. F1 SCORE & RECALL & PRECISION FOR EACH CLASS

|  | precision | recall | f1-score |
|---|---|---|---|
| **g0** | 1.00 | 1.00 | 1.00 |
| **g1** | 1.00 | 0.92 | 0.96 |
| **g2** | 1.00 | 1.00 | 1.00 |
| **g3** | 0.89 | 1.00 | 0.94 |
| **avg** | 0.97 | 0.98 | 0.97 |

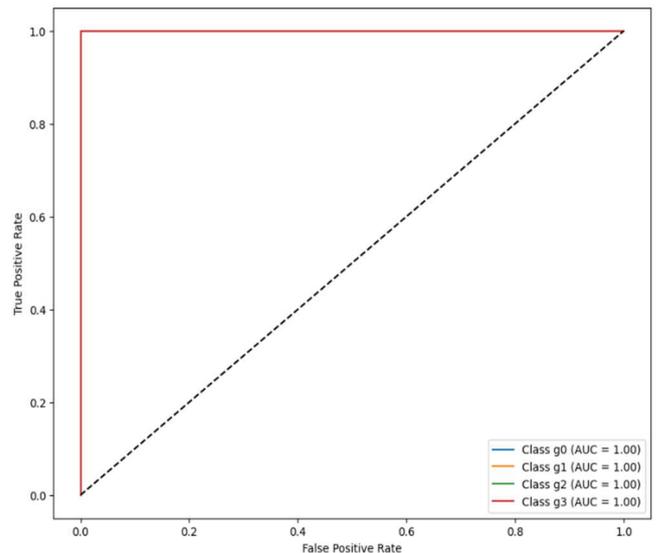

Fig. 6. ROC Curve for Each Class

## IV. CONCLUSIONS

The hardware system, featuring strategically positioned electrodes sequencing circuit, was pivotal in obtaining bioimpedance measurements from the knee by facilitating multi-electrode configurations for comprehensive data collection from various knee regions. The precise electrode arrangement enabled detailed impedance profiling, with consistent contact and placement ensuring accurate measurements. The use of cost-effective relays provided reliable switching without voltage drops, preserving the integrity of bioimpedance data by preventing significant signal interference. Complementing the hardware, the neural network model for classifying bioimpedance measurements demonstrated robust performance across all OA severity levels. The neural network model effectively utilized convolutional layers for feature extraction, dropout layers for regularization, and dense layers for classification, with early stopping and the Adam optimizer aiding in stable convergence and avoiding overfitting. This combined approach of bioimpedance measurements and neural networks offers a promising, non-invasive diagnostic method for knee OA, characterized by high accuracy and reliability. It holds potential for integration into clinical practices for early diagnosis and monitoring of OA progression and may extend to other future applications beyond OA to other musculoskeletal conditions.